\documentclass[sigconf]{acmart}
\pdfoutput=1
\usepackage{booktabs} % For formal tables

\usepackage{subfigure}
\usepackage{multirow}
\usepackage{algorithm}
\usepackage{algorithmic}
\usepackage{diagbox}
\begin{document}
\title{An Online Learning Based Path Selection for Multipath Video Telephony Service in Overlay}
\titlenote{Produces the permission block, and
  copyright information}

\author{Songyang Zhang}
\affiliation{
 \institution{School of Computer Science and Engineering, Northeastern University, China}
}
\email{sonyang.chang@foxmail.com}
\author{Weimin Lei}
\affiliation{
  \institution{School of Computer Science and Engineering, Northeastern University, China}
}
\email{leiweimin@ise.neu.edu.cn}
%\author{Wei Zhang}
%\affiliation{
%  \institution{School of Computer Science and Engineering, Northeastern University, China}
%}
%\email{zhangwei1@mail.neu.edu.cn}
%\author{Yunchong Guan}
%\affiliation{
%  \institution{School of Computer Science and Engineering, Northeastern University, China}
%}
%\email{y.c.guan@foxmail.com}
\begin{abstract}
Even real time video telephony services have been pervasively applied, providing satisfactory quality of experience to users is still a challenge task especially in wireless networks. Multipath transmission is a promising solution to improve video quality by aggregating bandwidth. In existing multipath transmission solutions, sender concurrently splits traffic on default routing paths and has no flexibility to select paths. When default paths fall into severe congestion and the available bandwidth decreases, sender has to decrease video quality by reducing resolution or encoding bitrate. Deploying relay servers in current infrastructure to form overlay network provides path diversity. An online learning approach based on multi-armed bandits is applied for path selection to harvest maximum profit. Further, a congestion control algorithm adapted from BBR is implemented to probe bandwidth and to avoid link congestion. To maintain throughput stability and fairness, a smaller probe up gain value is used and the cycle length in bandwidth probe phase is randomized. To reduce delay, the inflight packets will be reduced to match with the estimated bandwidth delay product in the probe down phase. Experiments are conducted to verify the effectiveness the proposed solution to improve throughput and quality in video communication service.
\end{abstract}


\maketitle

\section{Introduction}
According to a report \cite{cisco-report}, video traffic will occupy more than 80\% of all the traffic by the year 2022. The popular video telephony services push further increase of the video traffic. For WeChat alone, 410 million audio and video calls happened per day \cite{wechat-statistics}. Providing satisfactory of quality of experience for video streaming is still highly challenge \cite{Yu2014Can}, especially for the real time video telephony services \cite{Zhou2019Learning}. The best effort packets delivery mode of Internet does not provide any guarantee on quality of service. Bandwidth fluctuation, increased delay and packet loss would cause blurry images, video rendering process stalling, mosaic and skipped video frames. Such effects are quite annoying for users.

To improve user engagement \cite{Dobrian2011Understanding, Krishnan2013Video}, ensuring high quality of experience should be taken as priority for video service providers. Extensive solutions have been proposed to improve video streaming quality in current networks. In DASH (Dynamic Adaptive Streaming over HTTP) system, a same duration of video chunk is encoded with different quality. Clients make decision on different video chunks based on estimated bandwidth and buffered video length. The goal is to gain maximum video quality while reducing video play stall event. A playout buffer is deployed to absorb instantaneous throughput variation. 

The interactive video telephony services have stringent delay requirement and are allergic to bandwidth fluctuation. For conversational audio, ITU-T G.114 \cite{itu-delay} recommends for less than 150 milliseconds one-way delay for excellent  quality of experience, but delays between 150 and 400 milliseconds are still acceptable. Unlike DASH system, which applies a long buffer to insist bandwidth fluctuation, the video telephony applications usually deploy a congestion control algorithm at application layer to adapt the bitrate of video encoder to match with available bandwidth. For example, Rebera \cite{Kurdoglu2016Real} is a congestion control algorithm designed for WeChat International and GCC \cite{Carlucci2017Congestion} is the default congestion control algorithm deployed in WebRTC \footnote{https://webrtc.org/}. There are also other mechanisms e.g., negative acknowledgement (NACK), forward error correction (FEC) to combat packet loss for possible quality improvement in video telephony services.
Apart from these mentioned mechanisms, there are other possibilities to improve quality for video telephony. It is common for mobile devices equipped with multi-homed interfaces. The multipath transmission scheme can be exploited to improve video steaming quality by scheduling traffic over heterogeneous networks (4G and WiFi). Building overlay network \cite{Andersen2001Resilient} on existing Internet infrastructure is another option. With the development of cloud service, the geographically distributed datacenters are connected with managed backbone links, which provides an alternative to deploy relay servers to provide high performance service. TURN (Traversal Using Relays around NAT) servers are necessary to relay media traffic for real time video communication when direct communication between participating clients is not possible. These relay servers can be used to improve video call quality when default path falls into poor condition. By analyzing a dataset of 430 million calls from Skype, VIA \cite{Jiang2016Via} revisits the classic overlay networking techniques to reroute traffic for better call quality when the access link is not the bottleneck. An overlay network SD-RTN (software defined real time network) has been built by Agora.io \footnote{https://www.agora.io} in datacenters to provide intelligent routing service for video calls. Centralized servers could monitor the quality of overlay paths and recommend routings to users. It is claimed the SD-RTN can provide lower transmission delay and lower packet loss rate.

In this work, we combine overlay routing with multipath transmission scheme to improve quality of experience for video telephony service. In existing multipath transmission solutions \cite{Iyengar2006Concurrent, rfc6824}, sender has no flexibility to select paths. The video transmission quality is suffering when default paths fall into congestion and the available bandwidth is drastically reduced. Overlay network provides path diversity. An example is shown in Figure \ref{Fig:video-on-overlay}. With the help of relay nodes, the clients can choose routing path dynamically for each sub-flow to gain maximum profit.

There are some issues to be solved under such transmission scenario. How to select paths from several candidate paths with different path metrics to achieve maximum benefit? Without sending packets to a specific path, the path metrics remains unknown. Even the best path is chosen at present, it may become suboptimal after a short time. In such situation, insisting a chosen path will miss another higher quality path. To solve such dilemma, the multi-arm bandit (MAB) approach is applied for path selection, which is an efficient method to learn in uncertain environment. Tradeoff is made between exploiting the path with the best predicted performance and exploring alternatives for possible improvement. The second thing is to implement a congestion control algorithm for video transmission service. An algorithm inspired form BBR \cite{Cardwell2016BBR} to better adapt for real time video traffic is applied. The available bandwidth estimated by congestion control algorithm provides reference for path selection algorithm. 

The rest of this paper is organized as follows. In section 2, related works on multipath transmission and overlay network are briefly reviewed. The proposed multipath transmission solution is described in detail in section 3. Experiments are conducted and results are presented in Section 4. Section 5 is conclusion.
\section{Related work}
The bandwidth for streaming regular high definition video (720p, 30 frames per second) ranges from 2.5Mbps to 4Mbps. More bandwidth (3.5Mbps to 5Mbps) is required if streaming with higher frame rate (720p, 60fps). People have ever increasing demand on video quality. The video resolution has evolved from 480p to 720p to 1080p and even 4K ultra high definition video is preparing to stream over Internet. Even the internet infrastructure has made fast advance in recent years, the bandwidth for video streaming is still limited. For example, the Verizon 4G LTE wireless broadband can provide download speed between 5 and 12 Mbps and upload speed between 2 and 5 Mbps \cite{verizonwireless}. Due to dynamic of traffic load at the bottleneck, it would be hard to reach the maximum access upload rate for video telephony traffic. Especially when the call sessions traverse different ASes (Autonomous system), different countries, and the last mile links are usually not the bottleneck. In \cite{Dong2016Improving}, the authors test upload rates in TD-LTE networks. 42\% users achieve rates below 2Mbps. The sessions with average throughput below 400 kbps accounting for 40\% from Taobao-Live network measurement traces \cite{Zhou2019Learning}. 

Hence, it is an attractive feature to send packets simultaneously by use of multiple interfaces to gain higher throughput. Current proposed multipath transmission protocols CMP-SCTP \cite{Iyengar2006Concurrent} and MPTCP \cite{rfc6824} are mainly designed for bulk data transfer. There are several works focused on congestion control and packet scheduling in MPTCP context. Current proposed congestion control algorithms (LIA \cite{rfc6356}, OLIA \cite{Khalili2013MPTCP}, wVegas \cite{Cao2012Delay}) for MPTCP couple all subflows together to achieve friendliness if the subflows of multipath session traverse a same bottleneck. The scheduling algorithms DEMS \cite{Guo2017Accelerating}, STMS \cite{Shi2018STMS}, and QAware \cite{Shreedhar2018QAware} are mainly proposed to alleviate packets arriving out of order problem in MPTCP. Due to varying path property, packets may arrive out of order at receiver, which may get buffer resource fully occupied and would cause head of line blocking (HOL) \cite{Ferlin2016BLEST}. In kernel net stack, the buffer is pre-allocated. Once the buffer at receiver is fully occupied, the subsequent arriving packets will be dropped. HOL blocking leads suboptimal throughput in MPTCP since the rate back off after packet loss is not caused by congestion. 

It is inconvenient to apply MPTCP to enable multipath transmission since it requires OS kernel support at both sides. Some works explore multipath video streaming at application layer by initializing separate TCP connections. MSPlayer \cite{Chen2016MSPlayer} requests video different chunks from two different servers in DASH system by accessing WiFi and cellular networks. The bandwidth on each path is estimated by harmonic mean to mitigate the effect large outliers due to network throughput variation. Such multipath transmission strategy reduces start up delay and provides higher video quality. In \cite{Elgabli2018Optimized}, the video frames are encoded into different layers by scalable video coding (SVC) technologies. Layer selection is formed as an optimization problem based on throughput predication. 

These multipath transmission protocols that strictly guarantee in order delivery and reliable transmission by implementing Automatic Repeat reQuest (ARQ) are not appropriate for delay sensitive video telephony traffic. Once a packet is lost, the subsequent received packet would not be submitted to the upper layer until the lost packet is recovered by retransmission. Considerable latency is thus introduced.  Hence, the video telephony traffic is usually streaming over UDP and this work is no exception. MPRTP \cite{Singh2013MPRTP} is designed as an extension to RTP protocol for real time video communication. 

The implementation of overlay network to provide better service is not a new idea. In \cite {Ren20151Mbps}, overlay helpers are used to optimize global live video streaming. Experiments on real test beds indicate video transmission in overlay network can reduce scheduling delay and improve throughput. In \cite{Jones2017Overlay}, a solution on relay nodes placement in current network infrastructure to satisfy the full throughput region is proposed. The conclusion is that only a small fraction of controllable nodes is sufficient to achieve maximum throughput.
\section{Approach}
\begin{figure*}
\includegraphics[scale=0.65]{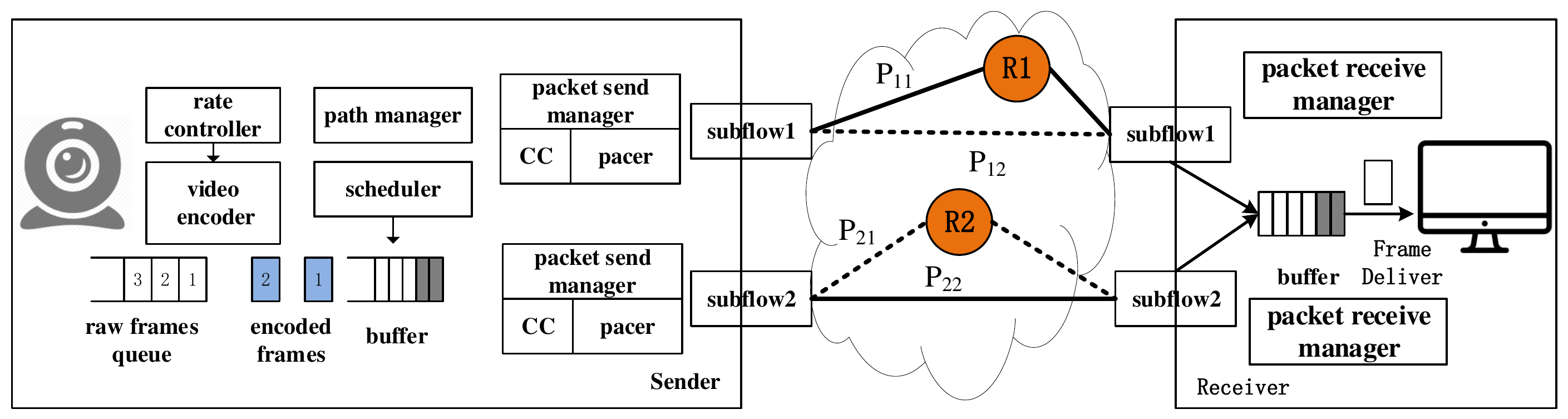}
\caption{Multipath Video Transmission of Multihomed Client in Overlay Network}
\label{Fig:video-on-overlay}
\end{figure*}

The multipath transmission system for video telephony in overlay network is illustrated in Figure \ref{Fig:video-on-overlay}. R denotes relay server. Sender distributes multimedia packets over different subflows. Each subflow can dynamically choose path (default path or overlay path formed by overlay nodes) for sending packets according to path quality. The relay nodes can be recommend by a centralized server based on historical statistics. In this work, we mainly focus on path selection to maximize video transmission quality, assuming relays have already been available for endpoints.
The modules in Figure \ref{Fig:video-on-overlay} at sender are responsible for packets transmission, path selection and bitrate adjustment on video encoder.
\subsection{Packets transmission}
For each path, there is a packet send manager implemented at sender side and a corresponding packet receive manager at receiver side. Only two pairs are shown in Figure \ref{Fig:video-on-overlay} due to space limitation. The estimated bandwidth and round trip delay can be got in packet send manager. At receiver sider, acknowledgement packet is sent to its peer at intervals when multimedia packets arrive. The congestion controller (CC) estimates available bandwidth based on feedback packets.
\subsubsection{Detail on congestion control algorithm}
The congestion control algorithm implemented in CC module is a modification from BBR \cite{Cardwell2016BBR} to adapt it for video telephony traffic transmission to maintain rate stability of video encoder. BBR is claimed as a congestion based rate control algorithm. Its goal is to get close to the optimal control point, in which flows achieve the maximum available bandwidth, minimum transmission delay and lowest packet loss rate. In BBR, when a sent packet is acknowledged, a roundtrip time sample and the inflight packet length ($\Delta delivered$) when the sent packet departs from sender can be got. A bandwidth estimation sample is calculated in Equation \ref{eq:bw-es}. The packet sending rate is $pacing\_rate$ in Equation \ref{eq:pacing}, which is the product of $pacing\_gain$ and maximum estimated throughput ($bw_{es}$) in 10 rounds.
\begin{equation}
\label{eq:bw-es}
bw=\frac{\Delta delivered}{\Delta t}
\end{equation}
\begin{figure}
\includegraphics[height=1in, width=3in]{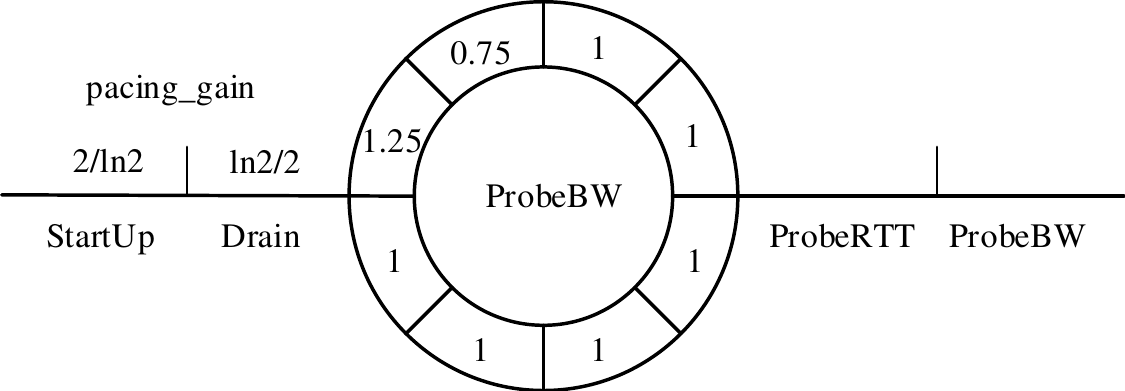}
\caption{Control states in BBR}
\label{Fig:bbr-state}
\end{figure}
\begin{equation}
\label{eq:pacing}
pacing\_rate=bw_{es}*pacing\_gain
\end{equation}

The four control states StartUp, Drain, ProbeBW, and ProbeRTT in BBR are shown in Figure \ref{Fig:bbr-state}.

In ProbeBW state, there are different $pacing\_gain$ values [1.25, 0.75, 1, 1, 1, 1, 1, 1] in 8 RTTs. The probe up phase with 1.25 gain is to increase inflight packets to probe extra bandwidth, and probe down phase with 0.75 gain is to get rid of excess queue. The congestion window is set as 2*BDP in ProbeBW to guarantee enough packets can be sent during probe up phase. If the minimal RTT is not sampled again within 10 seconds, the link seems falling into congestion, and it will enter into ProbeRTT state, and the congestion window is set as 4*MSS to get inflight packets totally drained from links. 

Even BBR achieves excellent performance in term of throughput for bulk data transmission in TCP, it also suffers from several drawbacks. Firstly, bandwidth allocation fairness can not be guaranteed in bottleneck link with shallow buffer. When multi flows share a bottleneck, flows overestimate bandwidth and overload the link. High packet loss rate is introduced. Secondly, BBR suffers from RTT unfairness issue and flavors towards flows with longer RTT. As indicated in \cite{Ma2017Fairness}, these strategic receivers could easily manipulate such drawback by sending delayed acknowledged packets to steal bandwidth. 

To implement BBR for real time video communication is an attempt in this work. In video telephony applications, video frames are captured at fixed intervals and the output bitrate of video encoder is stable in a short time span. In such situation, there may be not enough packets available in probe up phase with 1.25 gain for bandwidth probe. To maintain rate stability and avoid excess queueing delay, the pacing gain is 1.1 in probe up phase and the pacing gain is 0.85 in probe down phase. In our previous work \cite{Zhang2019Congestion}, a delay constraint BBR (Delay-BBR) algorithm is proposed in order to achieve lower transmission delay. Once delay signal exceeds a predefined threshold, Delay-BBR will actively reduce sending rate to drain inflight packets. In recent tests, we found it inherits the RTT unfairness property from BBR. In old version of BBR, the probe down phase holds for essentially $RTT_{min}$. In newly patch, the probe down phase holds the drain state until the inflight packets matching with estimated BDP. As verified in \cite{Zhang2019Evaluation}, the lower queue delay can be achieved and RTT unfairness issue is alleviated by this modification, but it results in rate fluctuation when flows competing for bandwidth. We found to randomize the gain cycle length from 2 to 8 can solve rate fluctuation problem. Such idea is applied in this work. This improved congestion control algorithm for video telephony traffic is named as RTC-BBR.

In RTC-BBR, the procedure to update pacing\_gain in ProbeBW is described in Algorithm \ref{alg:plusgain}. $CYCLE\_RAND$ is 7 and $kGainCycleLen$ is 8. $inflight$ denotes the length of sent packets that have not been acknowledged. When there is packet loss event, the probe up phase exists earlier (line 14 in Algorithm \ref{alg:plusgain}).
\begin{algorithm}[htb] 
\caption{UpdateGainCyclePhase} 
\label{alg:plusgain} 
\begin{algorithmic}[1]
\REQUIRE ~~\\ 
the timestamp (now), $inflight$, has\_loss
\STATE $elapsed\gets now-cycle\_mstamp\_$
\IF{$elapsed>cycle\_len\_*RTT_{min}$}
\STATE $cycle\_mstamp\_\gets now$
\STATE $cycle\_len\_\gets kGainCycleLen-rand()\%CYCLE\_RAND$
\STATE $pacing\_gain\gets 1.1$
\RETURN
\ENDIF
\IF{$pacing\_gain==1$}
\RETURN
\ENDIF
\IF{$pacing\_gain<1.0 \AND inflight\leq BDP$}
\STATE $pacing\_gain\gets 1$
\ENDIF
\IF{$elapsed>RTT_{min} \AND(inflight>1.1*BDP\ \OR\ has\_loss)$}
\STATE $pacing\_gain\gets 0.85$
\ENDIF
\end{algorithmic}
\end{algorithm}
\subsubsection{Pacer Module}
The connection will evenly send packets into network according to $pacing\_rate$. Traditionally, packets are seding in burst mode, which may lead packets queued at intermediate routers and introduce extra delay. 
When a packet A with length $L_a$ is sent out at time $a\_sent\_ts$, pacer calculates the next time (sent\_ts) that another packet is allowed to be sent out.
\begin{equation}
sent\_ts=a\_sent\_ts+\frac{L_a}{pacing\_rate}
\end{equation}
\subsubsection{Packet scheduling algorithm}
An encoded video frame will be packetized into segments with length smaller than maximum segment size. Apart from video payload, necessary information is tagged into header: frame index, frame captured timestamp, total segments of a frame and segment index. These segments are stored in buffer. The scheduler decides which subflow to send a specific packet in buffer.

The connection will maintain a queue in each subflow to record total length ($Q$) and packet offset for these scheduled packets. The time to send a new packet is depended on the pacer. When the time comes, the connection will find the packet in buffer with offset information and delivery it out.  

The goal of the implemented packets distribution algorithm in multipath context is to minimize packet expected arriving latency. The latency is composed of queue delay and transmission delay in Equation \eqref{eq:latency}. The information on smoothed roundtrip time $SRTT_s$ and available bandwidth $bw_{es}^s$ in subflow $s$ can be got from packet send manager. When a new RTT sample is avaialble, SRTT is updated according to Equation \eqref{eq:srtt}, and $delta$ is set as 0.85. When new packetized segments comes, it will be scheduled to the subflow with minimal delay in Equation \eqref{eq:latency}.

For packet belonging to a key frame, only when acknowledgment is received, it is evicted from buffer. For non-key frame packets, the sent packets will be cached at buffer at most 400 milliseconds. The lost packet that can be found in buffer will be retransmitted immediately over path with minimal transmission delay.
\begin{equation}
\label{eq:latency}
\lambda_s=\frac{SRTT_s}{2}+\frac{Q_s}{bw_{es}^s}
\end{equation}
\begin{equation}
\label{eq:srtt}
SRTT=(1-\delta)\times SRTT+\delta\times RTT
\end{equation}
\subsection{Path selection algorithm}
There are $k_i$ available paths for sub-flow $i$. The default path is included and the other $k_i-1$ paths are formed by relays.  The multi-homed caller in Figure \ref{Fig:video-on-overlay} can choose path from path set $P1=\{P_{11}, P_{12}\}$ for subflow1 and path set $P2=\{P_{21}, P_{22}\}$ is for subflow2. Transitional multipath transmission solutions only take the advantage of aggregating bandwidth. The path diversity provided by relays in overlay gives sender flexibility to choose routing paths to gain maximum profit. For example, when transmission quality of the default path P22 deteriorates (reduced bandwidth or insufferable transmission delay), subflow1 can switch to overlay path P21 for possible improvement. 

The rate distortion model in Equation \eqref{eq:distortion} is introduced in \cite{Stuhlmuller2000Analysis}. $R_e$ is the output bitrate of video encoder. $D_0$, $\theta$ and $R_0$ are parameters related to video sequence and codec. It clearly shows that lower encoded  image distortion can be achieved by increasing $R_e$. But for real time video transmission, $R_e$ should be matched with the reference rate $R$ set by Rate Controller. Hence, it is a role that path manager module can paly to choose high quality paths to reduce video distortion.
\begin{equation}
\label{eq:distortion}
D_e=\frac{\theta}{R_e-R_0}+D_0
\end{equation}

In each time slot, the path manager will choose a routing path for each sub-flow for packets transmission. $Q(i,j,t)$ denotes the transmission quality of path $j$ for subflow $i$ at time slot $t$. $a(i,j,t)$ indicates whether path $j$ is chosen at time slot $t$. The goal is to maximize video quality in the long run as Equation \ref{eq:max-quality} shows.
\begin{equation}
\label{eq:max-quality}
\begin{aligned}
&\max \frac{\sum_{t=1}^{T}\sum_{i}\sum_{j}Q(i,j,t)*a(i,j,t)}{T}\\
&\text{subject to}:a(i,j,t) \in \{0,1\}
\end{aligned}
\end{equation}

Due to lack of oracle perspective, to choose paths always provide maximum quality of experience to users is a hard task. Without sending packets to a specific path, the path metrics remains unknown. Even the best path is chosen at present, it may become suboptimal after a short time. The multi-armed bandit model is fit for such task. Because traffic load on each path is highly dynamic, path selection belongs to restless bandit problems. Each path can be regarded as an arm of a multiple slots machine. Once an arm ($j$) is pulled in time slot $t$, the path metrics (available bandwidth and delay) can be revealed from feedback packets and reward ($X(j,t)$) is generated. Always choosing an arm with maximum partially known properties may miss another arm with higher reward. Hence, tradeoff is made between exploration and exploitation. Exploration takes the risk by trying alternative choices to collection information on arms and exploitation just takes the best arm empirically with observed information.

The upper confidence bound (UCB) policy is applied here, which was analyzed in detail in \cite{Auer2002Finite}. In UCB, the upper bound reward of a specific arm is estimated by previous rewards and plus some uncertainty as Equation \eqref{eq:bound} shows.
\begin{equation}
\label{eq:bound}
I(j,t)=\overline X(j,t)+U(j,t)
\end{equation}
\begin{equation}
\label{eq:average}
\overline X(j,t)=\frac{\sum_{s=1}^t X(j,s)\boldsymbol 1(a(j,s))}{N(j,t)}
\end{equation}
\begin{equation}
\label{eq:U}
U(j,t)=B\sqrt{\frac{2log(T)}{N(j,t)}}
\end{equation}

$\overline X(k,t)$ is the mean of previous rewards. $N(j,t)$ counts the times that arm $j$ has been pulled from start. $U(j,t)$ is the upper confidence bound of rewards, which describes uncertainty of an arm. A common form for $U(j,t)$ is given in Equation \ref{eq:U}. Here, B is a tunable factor.

In each time slot, player selects the arm with maximum $I_{k,t}$. Equation \eqref{eq:bound} reflects well the tradeoff between exploration and exploitation. Based on this equation,  player decides whether to continue with the current arm or try alternatives. If the times to pull an arm are less, the bias on the estimated average mean is large. The term $U(k,t)$ encourages player to pull less selected arms to collect more samples to eliminate mean estimation bias. 

These involved parameters are explained here. Estimated bandwidth is used as reward for path manager to choose paths for sender. As indicated in \cite {Sutton2018Reinforcement}, the averaging methods in Equation \ref{eq:average} is appropriate for static bandit problems, in which reward probabilities remain unchanged over time. But available bandwidth on each path is nonstationary. When a path provides higher throughput for some time, similar throughput can be maintained in large possibility hereafter. It is reasonable to give more weight to recent rewards. The exponential smoothing filter is applied for such purpose and the term on empirical mean reward is redefined in Equation \eqref{eq:smooth-reward}. $\alpha$ is empirically set as 0.9. $bw$ is the newest estimated bandwidth from congestion controller when a path is exploited. $C$ denotes the paths to choose in each time slot, which is equal to the number of subflows.
\begin{equation}
\label{eq:smooth-reward}
\hat {Bw}(j)=(1-\alpha)\times\hat {Bw}(j)+\alpha\times bw
\end{equation}
\begin{equation}
\label{eq:upper-bound}
U(j,t)=Bw(j)\sqrt{\frac{2log(C\times T)}{N(j,t)}}
\end{equation}
\begin{equation}
\label{eq:path-selection}
a(i,j,t)=\mathop{\arg\max} \left\{\hat {Bw}(i,j)+Bw(j)\sqrt{\frac{2log(C\times T)}{N(i,j,t)}}\right\}
\end{equation}
The upper confidence bound term is given in Equation \eqref{eq:upper-bound}. The rule for path manager module to decide which path to use for subflow $i$ is given in Equation \eqref{eq:path-selection}. ${Bw(j)}$ is the maximum observed bandwidth during a monitor interval (kObservedTime=10 seconds). It can be interpreted that client expects to achieve the previous throughput by re-selecting this path. If the path is not exploited within the monitor interval, ${Bw(j)}$ is assigned with the maximum bandwidth sample observed in currently. It encourages client to choose this path again to collect bandwidth sample. When the path is chosen, ${Bw(j)}$ is updated with the new maximum estimated bandwidth during the path usage slot. The detail to update ${Bw(j)}$ is shown in Algorithm \ref{alg:bwsample}. Before each decision slot, the path manager will call Algorithm \ref{alg:deleteobsolete} to delete obsolete bandwidth samples.

The detail on path selection is show in Algorithm \ref{alg:ucb-pathselection}. When a path with the maximum reward (line 13) is chosen for a subflow, its pulled times counter $N$ will be updated (line 23). At session initial phase, all available paths will be exploited to collect throughput and path delay information. The initial value of N is 1. 
\begin{algorithm}[htb] 
\caption{OnNewBandwidthSample} 
\label{alg:bwsample} 
\begin{algorithmic}[1]
\REQUIRE ~~
$bw, now$
\STATE Push back ($bw, now$) to $bwSamples\_$
\IF{$samples\_==0$}
\STATE{$Bw\gets bw$}
\STATE{$maxBw\_\gets bw$}
\STATE{$\hat {Bw}=bw$}
\ELSE
\STATE{$\hat {Bw}=(1-\alpha)\times\hat {Bw}+\alpha\times bw$}
\ENDIF
\IF{$bw>maxBw\_$}
\STATE{$maxBw\_\gets bw$}
\ENDIF
\STATE{DeleteObsoleteSamples(now)}
\STATE{$samples\_\gets samples\_+1$}
\end{algorithmic}
\end{algorithm} 
\begin{algorithm}[htb] 
\caption{DeleteObsoleteSamples} 
\label{alg:deleteobsolete} 
\begin{algorithmic}[1]
\REQUIRE ~~
$now$
\WHILE {bwSamples\_.size()>1}
\STATE{$sample \gets bwSamples\_.front()$}
\IF{$now-sample.time>kObservedTime$}
\STATE{$bwSamples\_.pop\_front()$}
\ELSE
\STATE \textbf{break}
\ENDIF
\ENDWHILE
\STATE{$bw\gets 0$}
\FOR{each $sample \in bwSamples\_$}
\IF{$sample.bw>bw$}
\STATE{$bw\gets sample.bw$}
\ENDIF
\ENDFOR
\STATE{$Bw\gets bw$}
\IF{$bwSamples\_.size()==0$}
\STATE{$Bw\gets maxBw\_$}
\ENDIF
\end{algorithmic}
\end{algorithm}
\begin{algorithm}[htb] 
\caption{Path Selection Algorithm} 
\label{alg:ucb-pathselection} 
\begin{algorithmic}[1]
\REQUIRE ~~
$T$, time slots counter,\\
$subflows$, stores $flowid$ of all subflows,\\
$paths\_info$, stores $(id,flowid,Bw,\hat Bw, N)$ on all available paths
\ENSURE ~~
$exploited\_path$, the chosen path for subflow
\STATE $C=len(subflows)$
\STATE $P=len(paths\_info)$
\FOR{$i \in [0,C)$}
\STATE $flowid\gets subflows[i].id$
\STATE $X_{max} \gets 0$
\STATE $path\_id \gets -1$
\FOR{$j \in [0,P)$}
\IF{$paths\_info[j].flowid==flowid$}
\STATE $id\gets paths\_info[j].id$
\STATE $\hat {Bw}\gets paths\_info[j].\hat {Bw}$
\STATE $Bw\gets paths\_info[j].Bw$
\STATE $N\gets paths\_info[j].N$
\STATE $X \gets \hat {Bw}+ Bw\sqrt{\frac{2log(C\times T)}{N}}$
\IF {$X > X_{max}$}
\STATE $X_{max}\gets X$
\STATE $path\_id=id$
\ENDIF
\ENDIF
\ENDFOR
\STATE $subflows[i].exploited\_path=path\_id$
\FOR{$j \in [0,P)$}
\IF{paths\_info[j].id==path\_id}
\STATE $paths\_info[j].N\gets paths\_info[j].N+1$
\ENDIF
\ENDFOR
\ENDFOR
\STATE $T\gets T+1$
\end{algorithmic}
\end{algorithm} 
\subsection{Reference bitrate adjustment for encoder}
Each captured video frame is firstly delivered to raw frame queue waiting to be processed by encoder. The packets of encoded images will be put into buffer. 

Rate controller adjusts the output bitrate of video encoder with reference rate ($R=\sum_{s\in S} bw_s$). $bw_s$ is got from congestion controllers of currently exploited routing path in sublfow $s$. The reference bitrate of encoder is reconfigured every 50 milliseconds.

We observed in experiments that video encoder can not immediately generate bitrate matching with the target rate. Especially when available bitrate is decreased, encoder takes about 1 second to output encoded images approximating the target bitrate, which is also revealed in \cite{Zhou2019Learning}. In such situation, the length of buffer will increase. In order to assure low frame delivery delay, it is common to drop some raw frames when delay exceeds certain threshold. A method to reduce bitrate by reducing frame rate.

The delay ($d_s$) before packets of an captured frame can be delivered at sender is composed by three parts: delay ($d_q$) at raw frame queue, encoding delay ($d_{en}$) and buffer delay ($d_b$). When a raw frame is sent to encoder, a sample of encoding delay $d_{en}(i)$ is got after the frame is encoded. The exponential smooth filter is applied again in Equation \eqref{eq:smooth-en} to update the estimation on image encoding delay. 

When a raw frame is dequeued, it will be dropped if the delay in Equation \eqref{eq:drop-delay} exceeds 400 milliseconds. Here, $\lambda_{min}=\min_{s\in S} \lambda_s$.
\begin{equation}
\label{eq:smooth-en}
\hat d_{en}=(1-\alpha)\times\hat d_{en}+\alpha\times d_{en}(i)
\end{equation}
\begin{equation}
\label{eq:drop-delay}
d=d_q+\hat d_{en}+\lambda_{min}
\end{equation}
\subsection{Transmission protocol}
For low latency consideration, packets in RTC applications are usually sent over UDP with partially reliability. To evaluate the performance of the proposed multipath transmission solution in this work, a transmission protocol is implemented. It is referenced from QUIC \footnote{https://www.chromium.org/quic} and only three frames (STREAM, STOP WAITING and ACK) are applied. Each sent packet is allocated with a unique packet number and the ACK frame is sent to its peer to notify received packets information. Receiver sorts received packets by offset number extracted from STREAM frame. The STOP WAITING frame notifies the peer to stop waiting these packets with packet number under the notified number. Such designation leaves sender flexibility to decide whether to retransmit lost STREAM frames according to packet importance and delay. Each retransmitted packet is allocated a new packet number.
\section{Evaluation}
All experiments involved in this part are running on ns3.26. The collected simulation data in this work is publicly available at \footnote{https://pan.baidu.com/s/1Y3ZTQxBFpyjCmjTuYCKNzA:yty3}
\begin{figure}
\includegraphics[height=1.2in, width=3in]{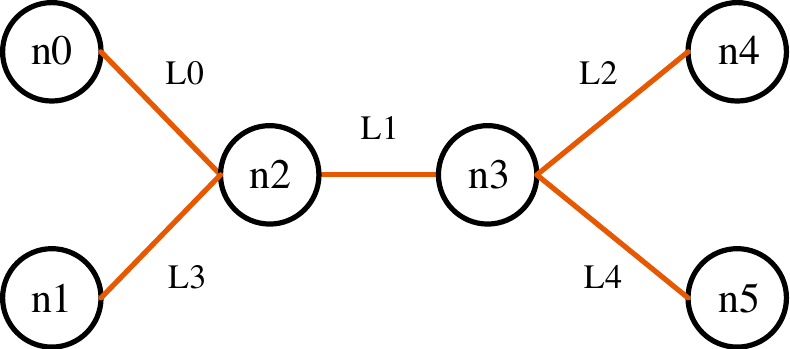}
\caption{Dumbbell topology}
\label{Fig:dumbbell}
\end{figure}
\subsection{Evaluation on congestion control algorithm}
\begin{table}[]
\centering
\caption{Link configuration on  L1}
\label{tab:l1}
\begin{tabular}{|c|c|c|c|}
\hline
Case & Bandwidth & Propagation Delay & Queue Length \\ \hline
1    & 3Mbps     & 50ms          & 3Mbps*100ms  \\ \hline
2    & 3Mbps     & 50ms          & 3Mbps*150ms  \\ \hline
3    & 3Mbps     & 50ms          & 3Mbps*200ms  \\ \hline
4    & 5Mbps     & 50ms          & 5Mbps*100ms  \\ \hline
5    & 5Mbps     & 50ms          & 5Mbps*150ms  \\ \hline
6    & 5Mbps     & 50ms          & 5Mbps*200ms  \\ \hline
7    & 6Mbps     & 50ms          & 6Mbps*150ms  \\ \hline
8    & 6Mbps     & 50ms          & 6Mbps*200ms  \\ \hline
9    & 8Mbps     & 50ms          & 8Mbps*150ms  \\ \hline
10   & 8Mbps     & 50ms          & 8Mbps*200ms  \\ \hline
11   & 10Mbps    & 50ms          & 10Mbps*150ms \\ \hline
12   & 10Mbps    & 50ms          & 10Mbps*200ms \\ \hline
\end{tabular}
\end{table}
\begin{figure}
\centering
\includegraphics[width=3in]{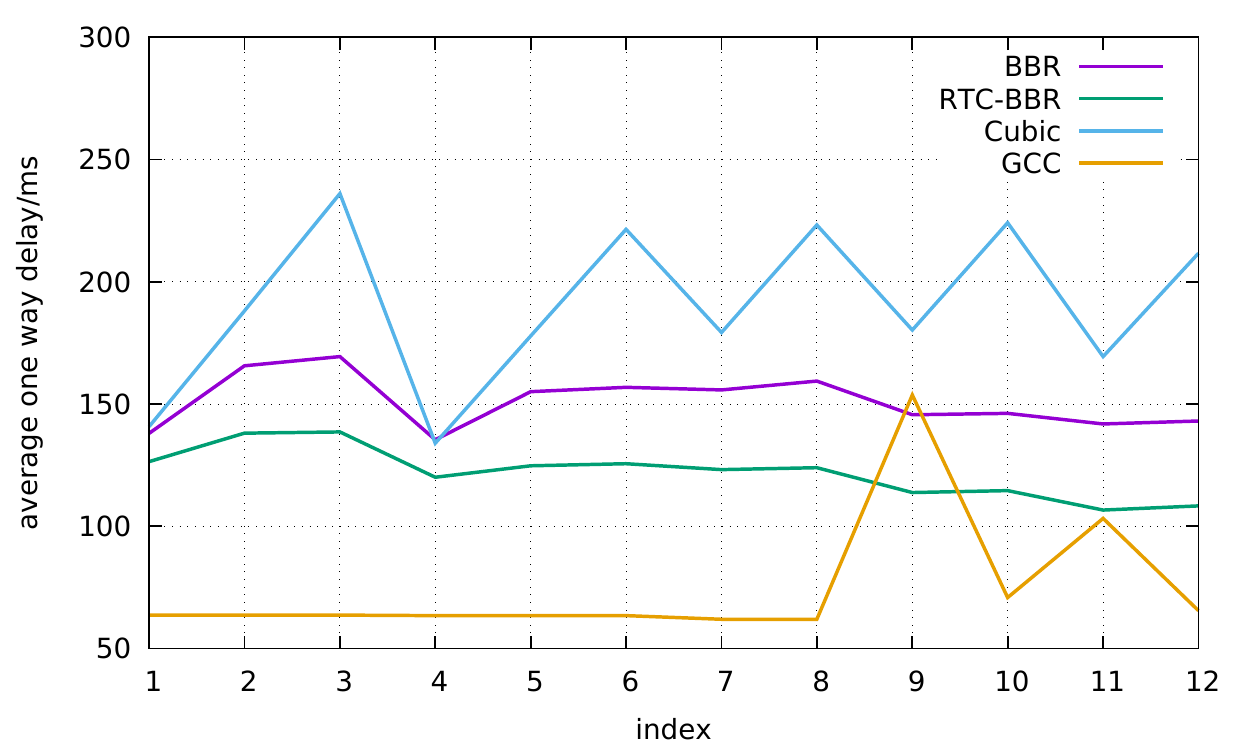}
\caption{Average one way transmission delay}
\label{Fig:owd-comapre}
\end{figure}
\begin{figure}
\centering
\includegraphics[width=3in]{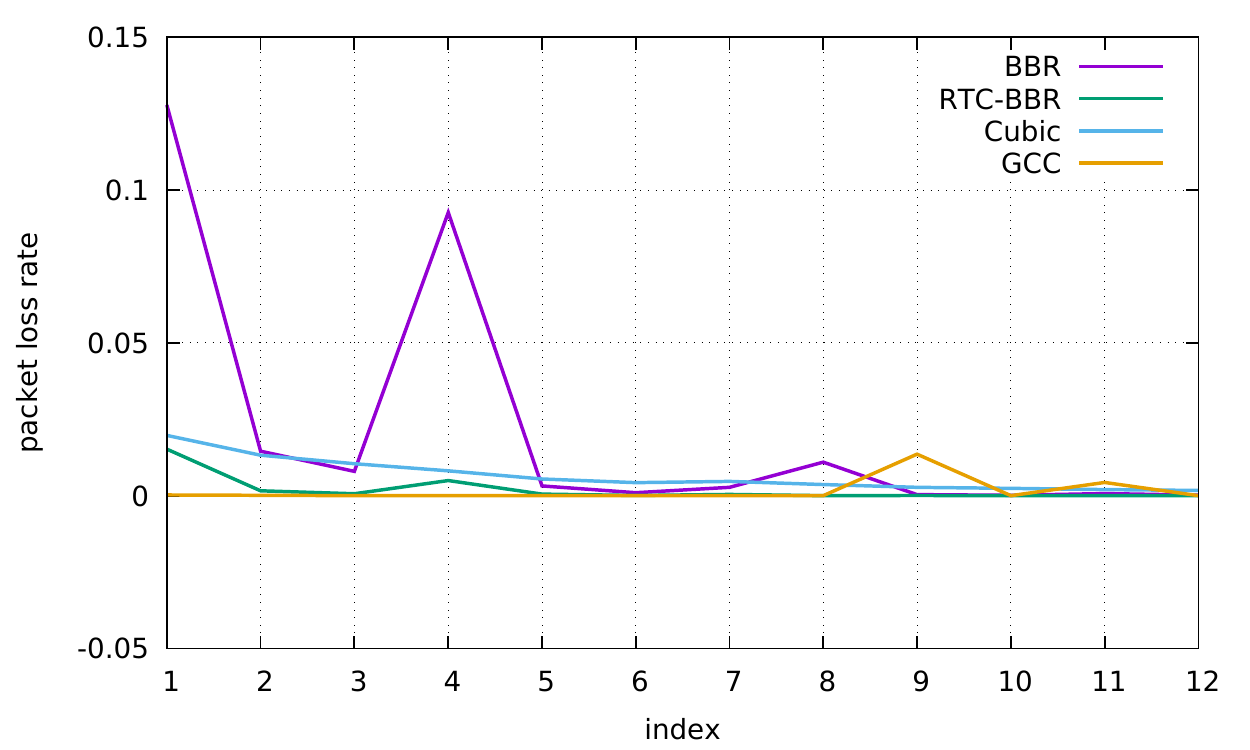}
\caption{Packet loss rate}
\label{Fig:loss-comapre}
\end{figure}
\begin{figure*}
\centering
\subfigure[]{\includegraphics[width=2in]{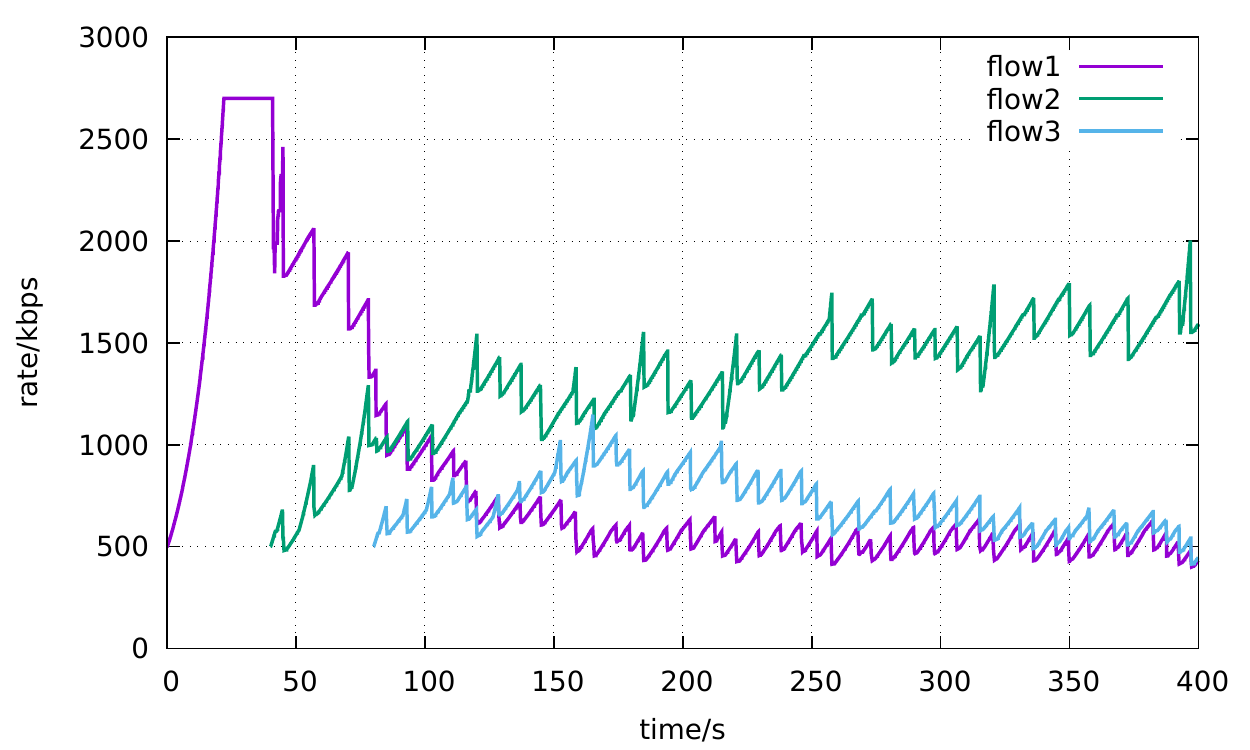}}
\subfigure[]{\includegraphics[width=2in]{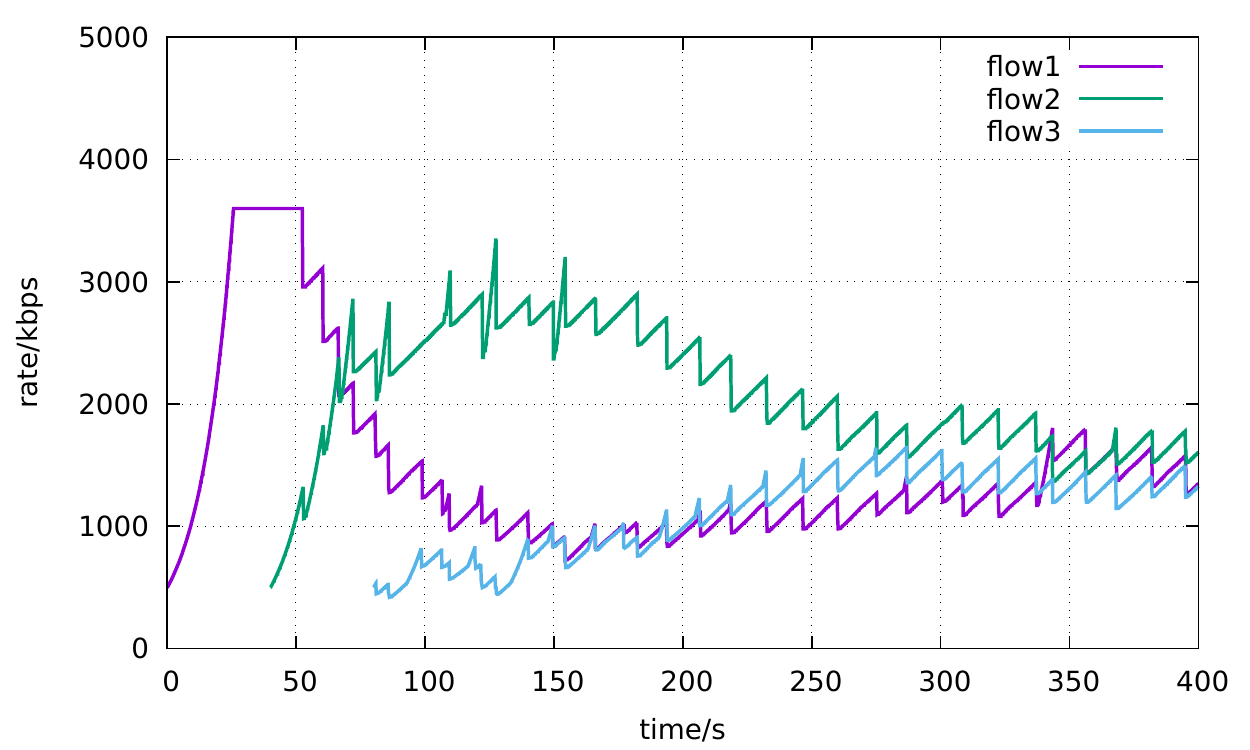}}
\subfigure[]{\includegraphics[width=2in]{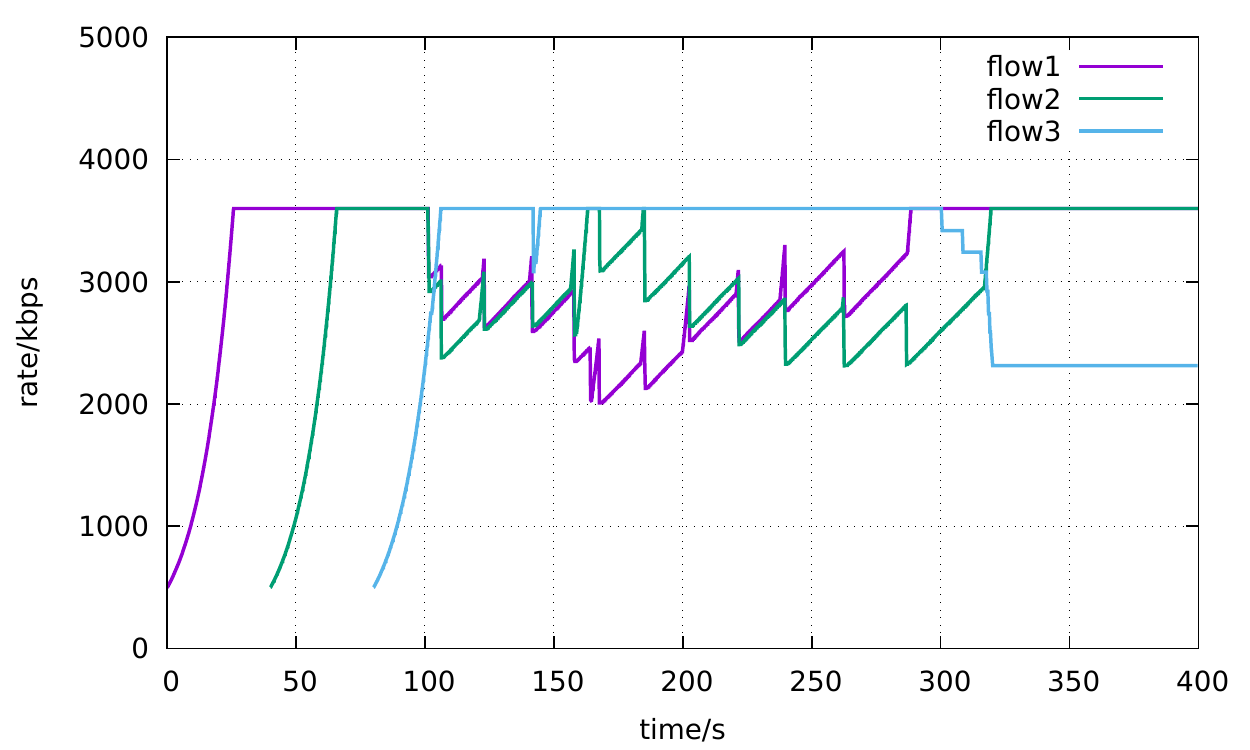}}
\caption{Rate dynamics of GCC flows. (a) C1.  (b) C5. (c) C11.}
\label{Fig:gcc}
\end{figure*} 
\begin{figure*}
\centering
\subfigure[]{\includegraphics[width=2in]{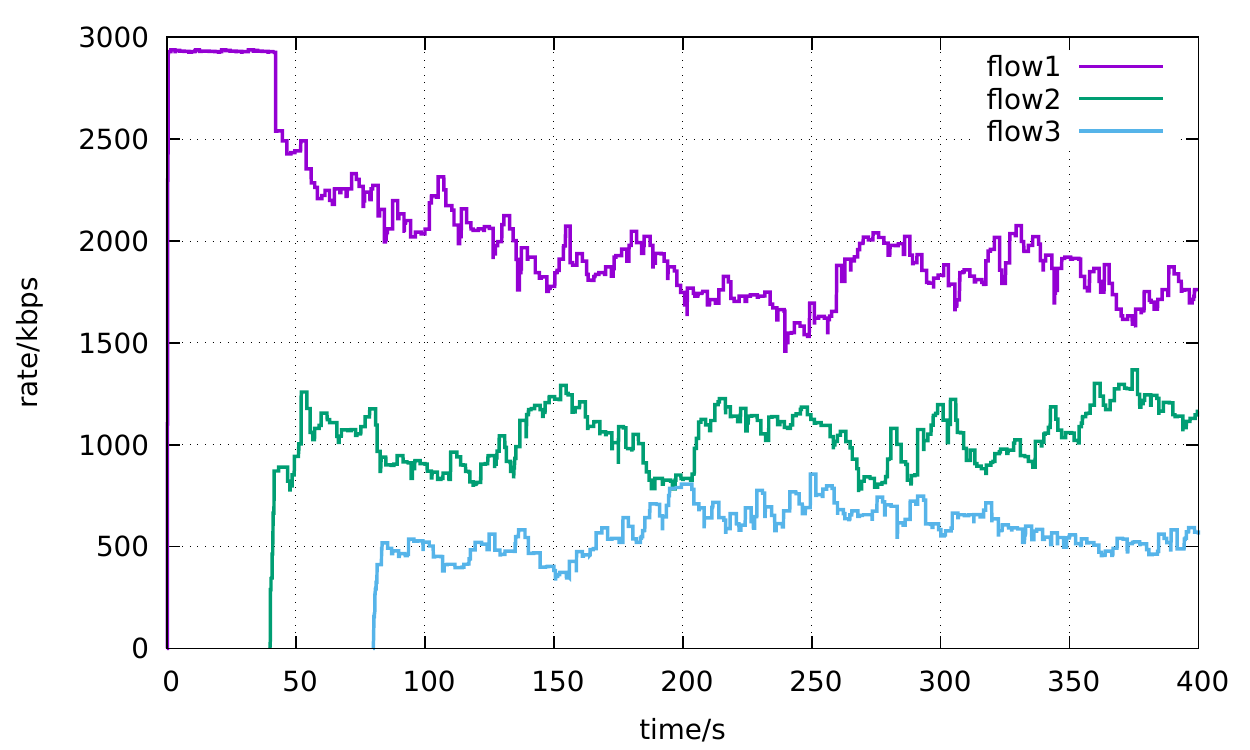}}
\subfigure[]{\includegraphics[width=2in]{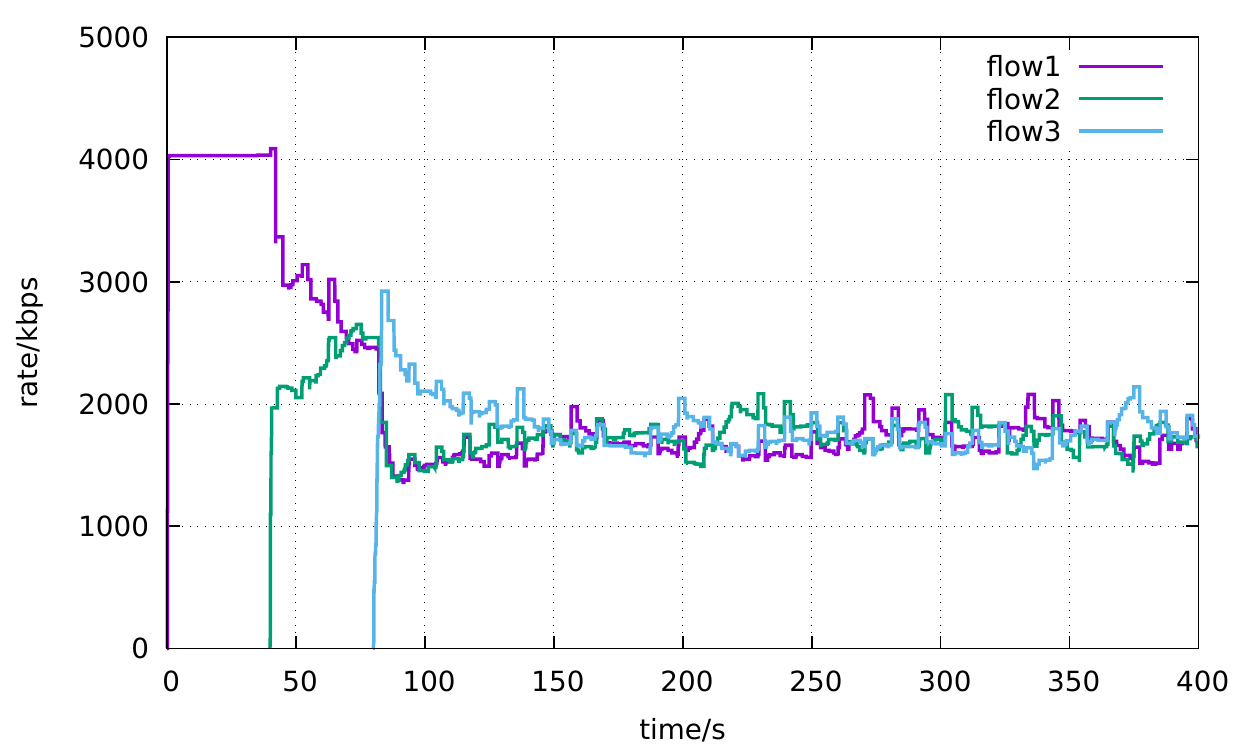}}
\subfigure[]{\includegraphics[width=2in]{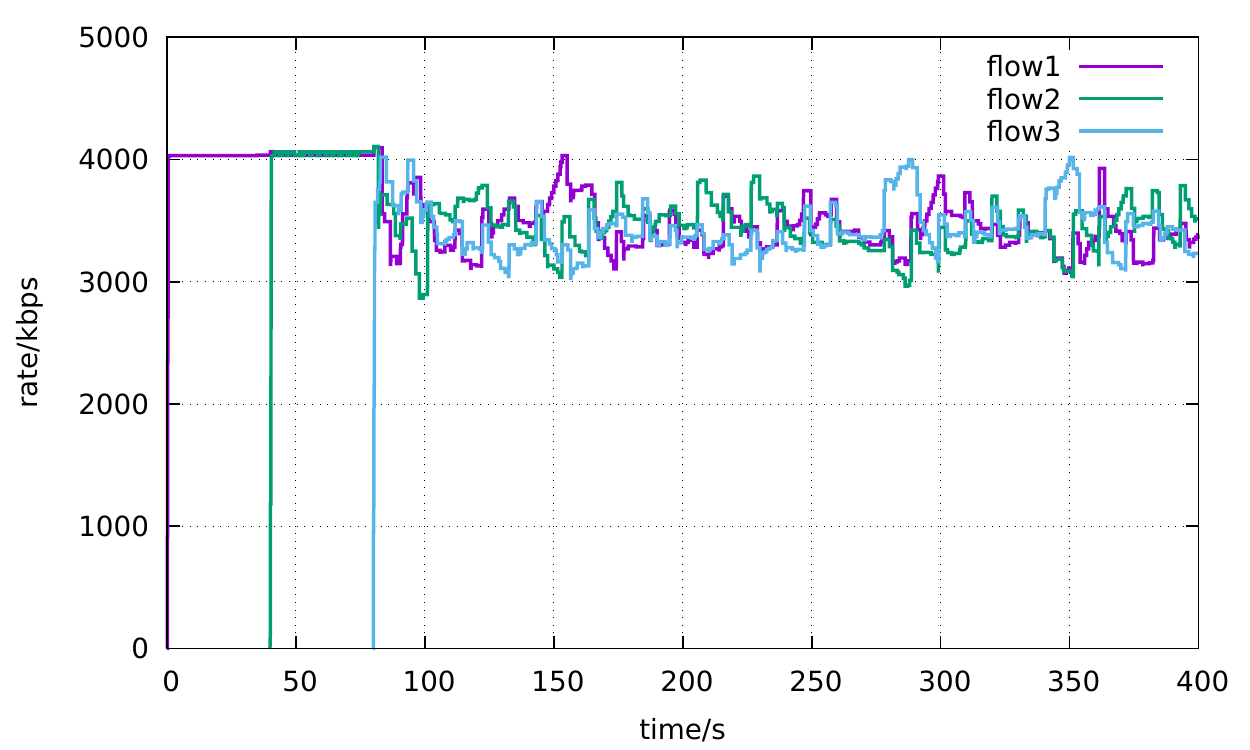}}
\caption{Rate dynamics of BBR flows. (a) C1.  (b) C5. (c) C11.}
\label{Fig:bbr}
\end{figure*} 
\begin{figure*}
\centering
\subfigure[]{\includegraphics[width=2in]{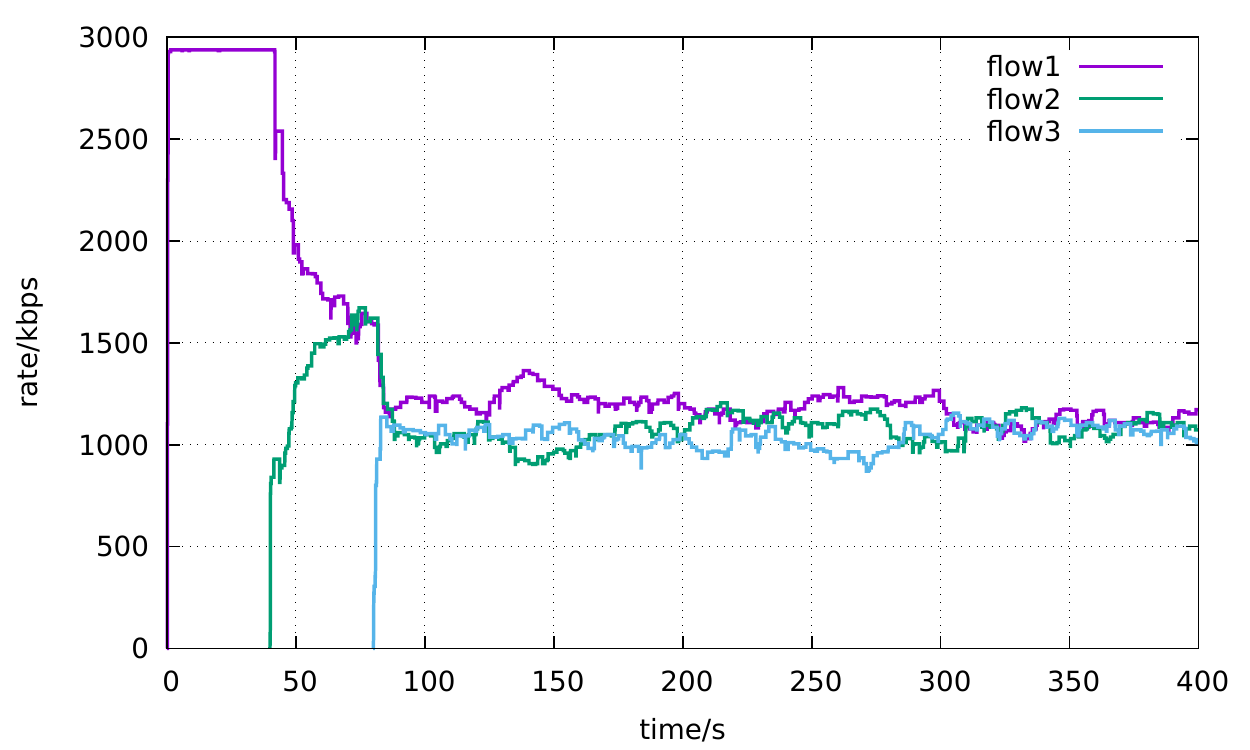}}
\subfigure[]{\includegraphics[width=2in]{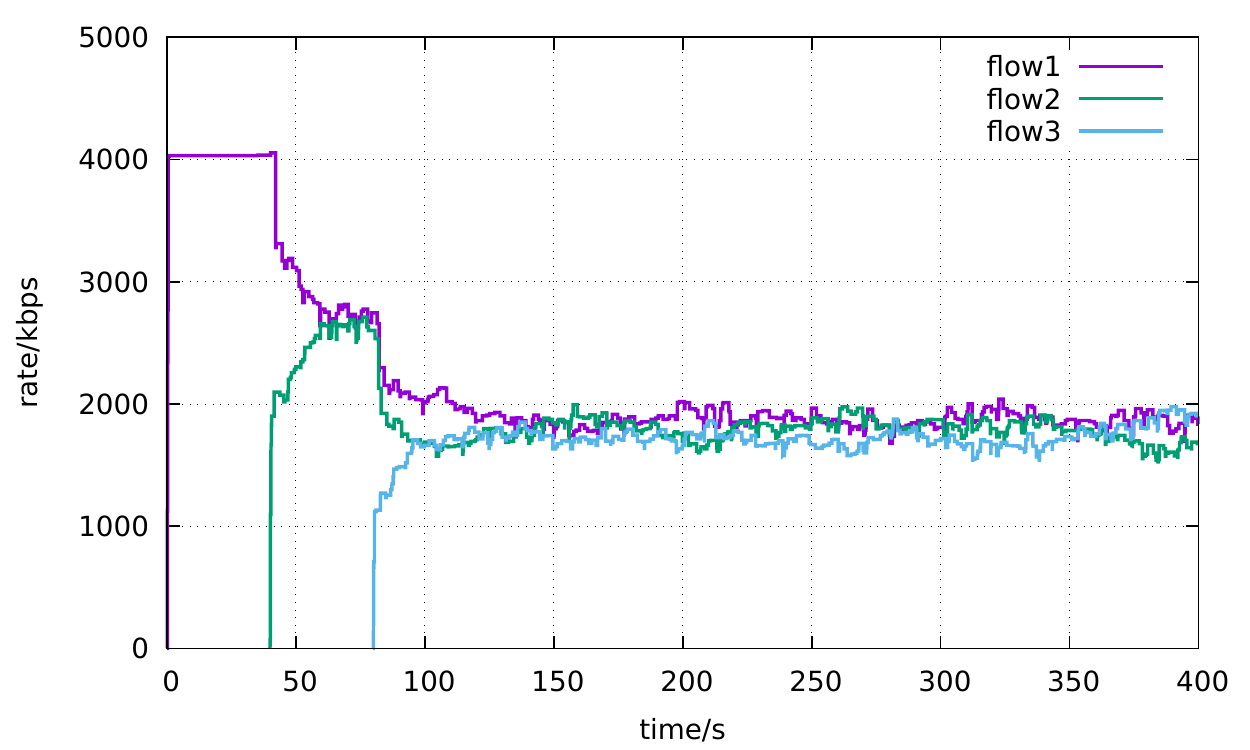}}
\subfigure[]{\includegraphics[width=2in]{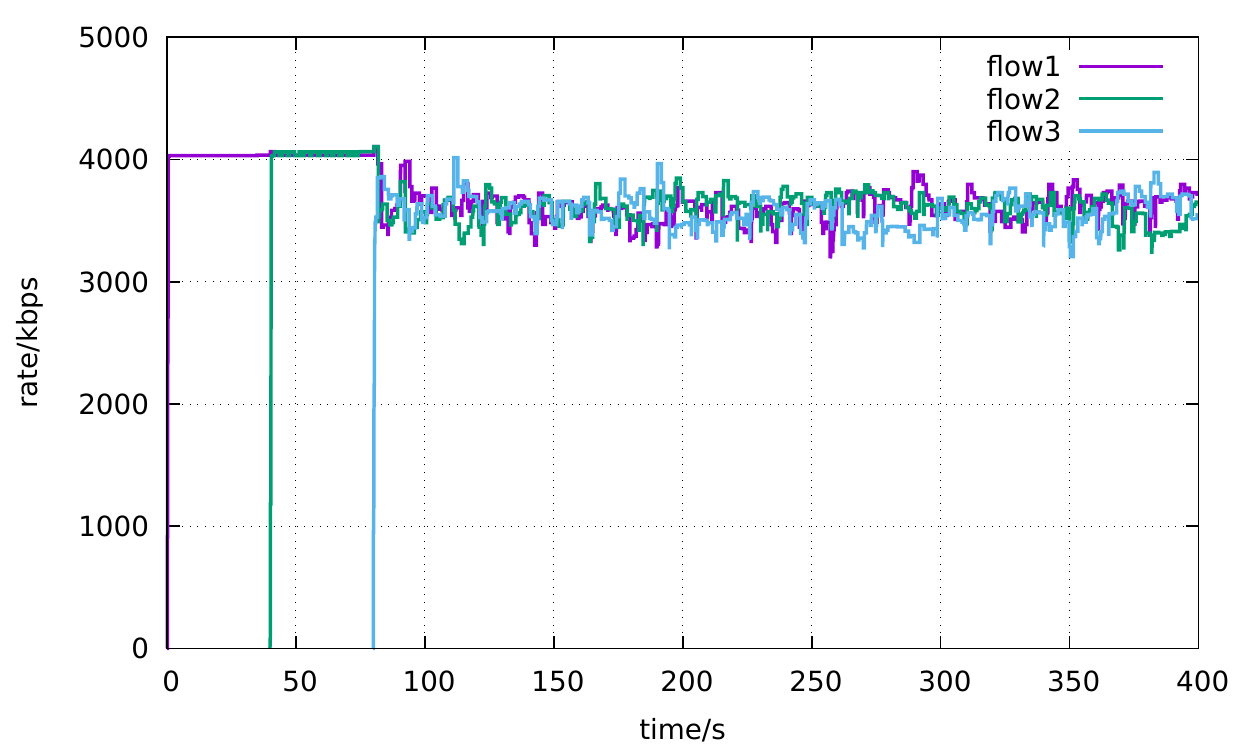}}
\caption{Rate dynamics of RTC-BBR flows. (a) C1.  (b) C5. (c) C11.}
\label{Fig:rtcbbr}
\end{figure*} 

To test the performance of RTC-BBR algorithm, a dumbbell topology in Figure \ref{Fig:dumbbell} is built on ns3. Apart from BBR, Cubic \cite{Ha2008CUBIC} and GCC are also evaluated to work as benchmarks. The configuration on link $L1$ is given in Table \ref{tab:l1}. The parameters are link capacity (in unit of Mbps), one way propagation delay (in unit of millisecond) and queue length in routers. Droptail queue management is implemented in these routers. Once the buffered packets exceed the maximum queue length, extra incoming packets will be dropped. There are 12 experiments in total.

In each test, three flows following the same congestion control algorithm start to send packets at different time. The second flow is started at 40s and the third flow is initialized at 80s. Each simulation process lasts 400 seconds. For real time video traffic, packets generating rate can not increase without bound. The maximum transmission rate is set as 4Mbps. At sender sider, when a new packet is sent, the estimated bandwidth of the congestion controller is traced. The sent time of each packet is tagged into the packet object in ns3, which could be used by receiver to calculate one way transmission delay. One way transmission delay is an indicator to buffer occupation in routers. 

According to the collected data, the average transmission delay and packet loss rate of all flows in each test are calculated. The results of are given in Figure \ref{Fig:owd-comapre} and Figure \ref{Fig:loss-comapre}.
From Figure \ref{Fig:owd-comapre}, RTC-BBR flows achieve lower transmission delay than BBR flows. Because GCC makes earlier action when bottleneck is in congestion by exploiting one way delay gradient to infer congestion, GCC flows achieve the lowest delay in most cases. Cubic flows have the largest delay. Cubic takes packet loss as congestion signal. It keeps increase its rate when no packet loss event happens, which would get the buffer in bottleneck fully occupied. Such rate control behavior is the root cause of Bufferbloat \cite{Gettys2011Bufferbloat}, which is a new form of congestion collapse in today's networks.

The rate dynamic of each flow in Case 1, Case 5 and Case 11 are given here for further analysis. When bottleneck link is configured with shallow buffer (less than 1.5*BDP), the bandwidth allocation fairness can not be guaranteed in BBR and an example is shown in Figure \ref{Fig:bbr}(a) in experiment 1. Most bandwidth is occupied by the first flow. There are also considerable packet loss rates (about 12\% in Case 1 and 9\% in Case 4) in Figure \ref{Fig:loss-comapre}. BBR algorithm tends to overestimate available bandwidth when flows share a bottleneck, which would lead bottleneck link overloaded. It is the reason behind such high packet loss rate. RTC-BBR flows show better performance in term of bandwidth allocation fairness under the same link configuration as shown in Figure \ref{Fig:rtcbbr}(a). At stable phase, the rate dynamics of RTC-BBR flows are more stable (Figure \ref{Fig:rtcbbr}(b) and Figure \ref{Fig:rtcbbr}(c)) when compared with BBR flows (Figure \ref{Fig:bbr}(b) and Figure \ref{Fig:bbr}(c)).

Even though GCC achieves lowest transmission delay and lowest packet loss rate in most cases, it suffers from several drawbacks. The bandwidth allocation unfairness is also observed in Figure \ref{Fig:gcc}(a) and Figure \ref{Fig:gcc}(c). Most importantly, GCC takes quite long time to reach the maximum throughput when there is extra bandwidth available. The GCC flow1 shown in Figure \ref{Fig:gcc}(a) takes about 22 seconds to achieve the maximum rate 2.7Mbps, while RTC-BBR flow1 shown in Figure \ref{Fig:rtcbbr}(a) takes about 0.56 seconds to achieve the maximum rate 2.9Mbps. The ability to make fast probe to the maximum available bandwidth can get the bandwidth resource highly utilized, which is an attractive feature of BBR-like algorithms.
\begin{table}[]
\centering
\caption{Configuration of links to test rtt unfairness}
\label{tab:all-config}
\scalebox{0.85}{
\begin{tabular}{|c|c|c|c|c|c|}
\hline
\multirow{2}{*}{Case} & L0            & L1           & L2            & L3           & L4           \\ \cline{2-6} 
                      & \multicolumn{5}{c|}{(BW, OWD, Q)}                                          \\ \hline
1                     & (10, 10, 200) & (4, 10, 200) & (10, 10, 200) & (10, 20, 200) & (10, 30, 200) \\ \hline
2                     & (10, 10, 200) & (4, 10, 200) & (10, 10, 200) & (10, 10, 200) & (10, 30, 200) \\ \hline
3                     & (10, 20, 200) & (4, 10, 200) & (10, 10, 200) & (10, 10, 200) & (10, 30, 200) \\ \hline
\end{tabular}}
\end{table}
\begin{table}[]
\centering
\caption{Calculated results in RTT unfairness simulation}
\label{tab:rtt-result}
\scalebox{0.8}{
\begin{tabular}{|c|c|c|c|}
\hline
\multirow{2}{*}{\diagbox{Algo}{Case}} & 1          & 2          & 3         \\ \cline{2-4} 
                  & \multicolumn{3}{c|}{($\overline {x_1}$, $\overline {x_2}$, R)} \\ \hline
BBR               &(634, 3155, 4.98)&(837, 2972, 3.55)&(1338, 2471, 1.85)\\ \hline
RTC-BBR            &(1792, 2026, 1.13)&(1835, 1988, 1.08)&(1921, 1880, 1.02)\\ \hline
\end{tabular}}
\end{table}
%\begin{figure}[!htb]
%\begin{tabular}{cc}
%\subfigure[BBR]{
%\begin{minipage}[t]{0.5\linewidth}
%    \includegraphics[width = 1\linewidth]{bbr-unfair-bw.pdf}
%\end{minipage}}
%\subfigure[RTC-BBR]{
%\begin{minipage}[t]{0.5\linewidth}
%    \includegraphics[width = 1\linewidth]{rtcbbr-unfair-bw.pdf}  
%\end{minipage}}
%\end{tabular}
%\caption{Rate dynamics of flows in case 3}
%\label{Fig:rtt-unfair-example} 
%\end{figure}
\begin{figure}
\centering
\includegraphics[width=3in]{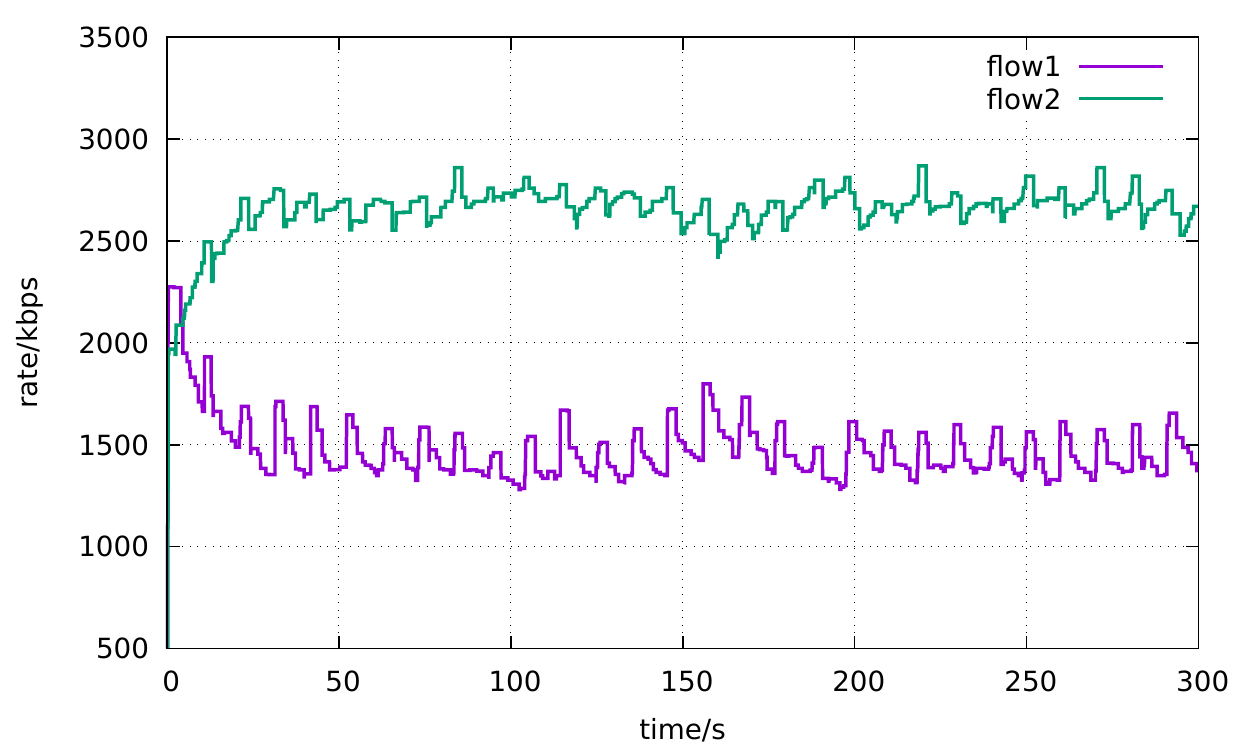}
\caption{RTT unfairness issue of BBR in case 3}
\label{Fig:bbr-rtt-unfair}
\end{figure}
\begin{figure}
\centering
\includegraphics[width=3in]{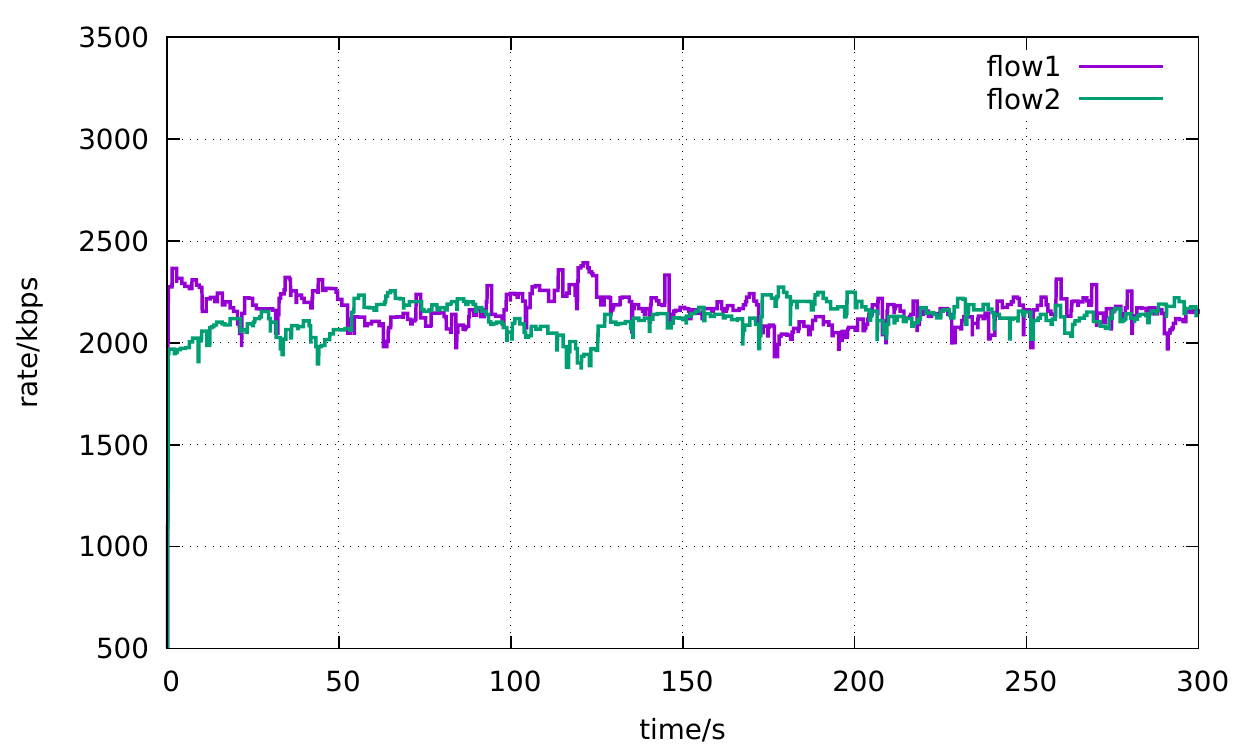}
\caption{RTT fairness improvement of RTC-BBR in case 3}
\label{Fig:rtcbbr-rtt-unfair}
\end{figure}

As indicated in \cite{Hock2017Experimental, Ma2017Fairness}, BBR flavors towards flows with longer round trip delay. The dumbbell topology is applied to verify whether RTC-BBR algorithm suffers from the same RTT unfairness problem. The configuration of all links is shown in Table \ref{tab:all-config}. $Q$ is in unit of milliseconds and the buffer length of each link will be configure as $BW*Q$. Three experiments are designed. In each case, two flows are running at the same time and are constrained under a same congestion control algorithm(BBR or RTC-BBR). flow1 starts from n0 to destination n4 (path1) and flow2 sends packets through path2 (n1 to n5). The running time of each experiment lasts 300 seconds. The average throughput is calculated as Equation \ref{eq:average-rate}. $bytes$ is the length of all received packets at receiver. The throughput ratio defined in Equation \ref{eq:ratio} of flow2 and flow2 is calculated. The results are given in Table \ref{tab:rtt-result} and The average rates of the two RTC-BBR flows are quite close in each test. BBR flow2 can approach higher throughput in three tests. The rate dynamics of case 3 are given in Figure \ref{Fig:bbr-rtt-unfair} and \ref{Fig:rtcbbr-rtt-unfair}. RTC-BBR algorithm can obviously alleviate the RTT unfairness issue.
\begin{equation}
\label{eq:average-rate}
\overline x=\frac{bytes}{duration}
\end{equation}
\begin{equation}
\label{eq:ratio}
R=\frac{\overline {x_2}}{\overline {x_1}}
\end{equation}
\begin{figure}
\includegraphics[height=2in, width=3in]{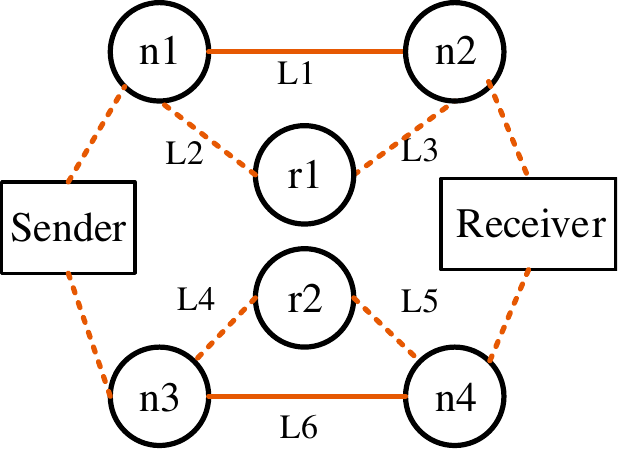}
\caption{Mulitpath topology}
\label{Fig:multipath}
\end{figure}
\begin{figure}
\centering
\includegraphics[width=3in]{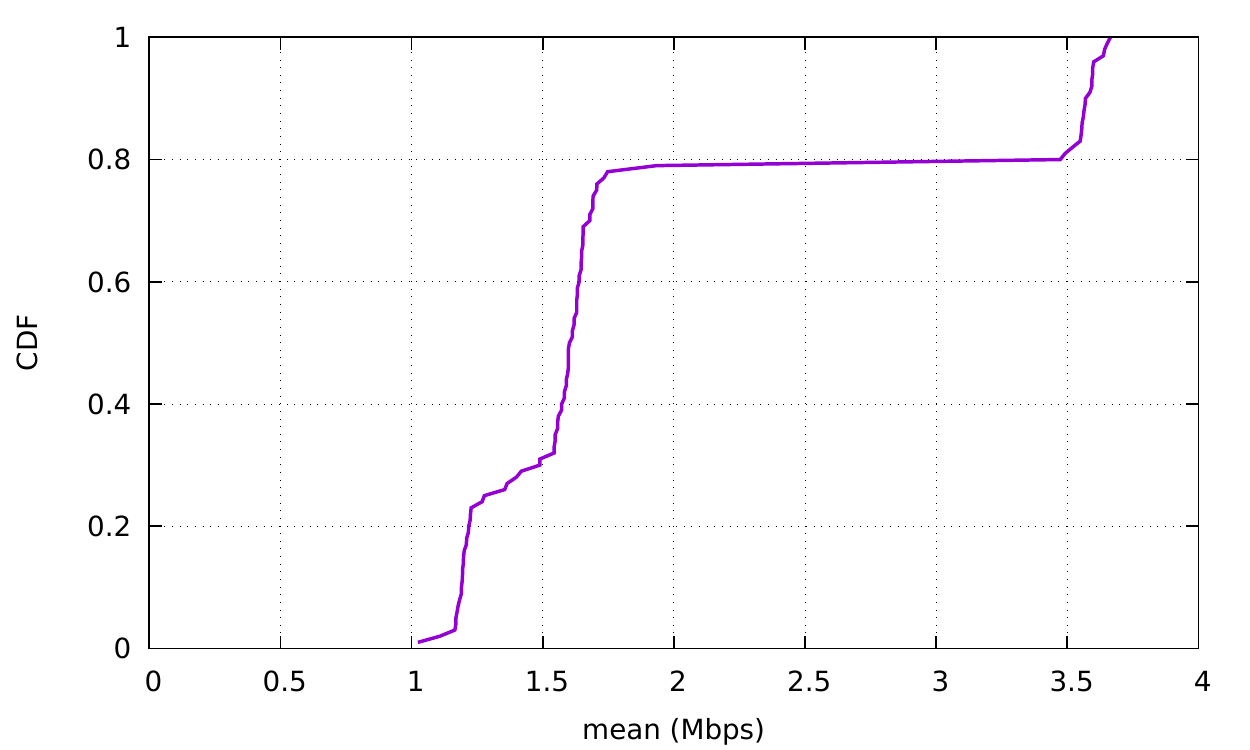}
\caption{Cumulative average rate distribution of traced dataset}
\label{Fig:trace-rate-cdf}
\end{figure}
\begin{figure}
\centering
\includegraphics[width=3in]{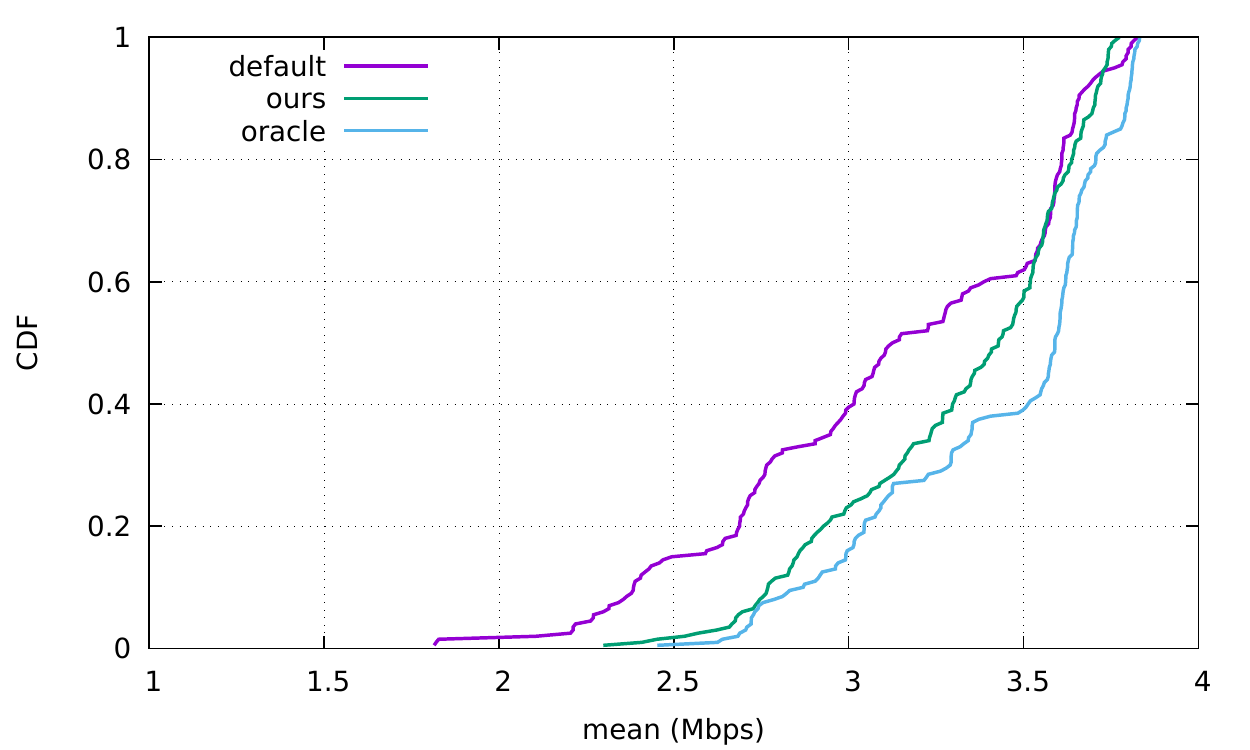}
\caption{Cumulative distribution of average throughput}
\label{Fig:rate-cdf}
\end{figure}
\begin{figure}
\centering
\includegraphics[width=3in]{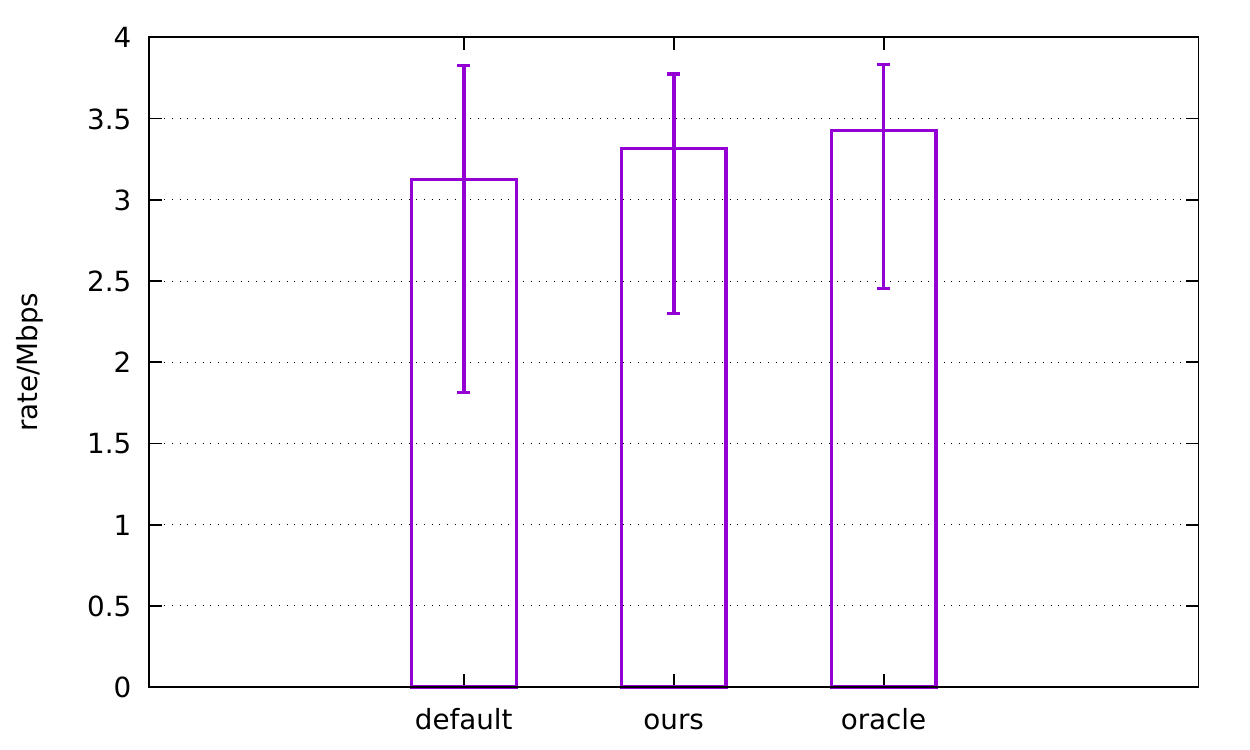}
\caption{Error bar on average rate of all tests}
\label{Fig:rate-error}
\end{figure}
\begin{figure}
\centering
\includegraphics[width=3in]{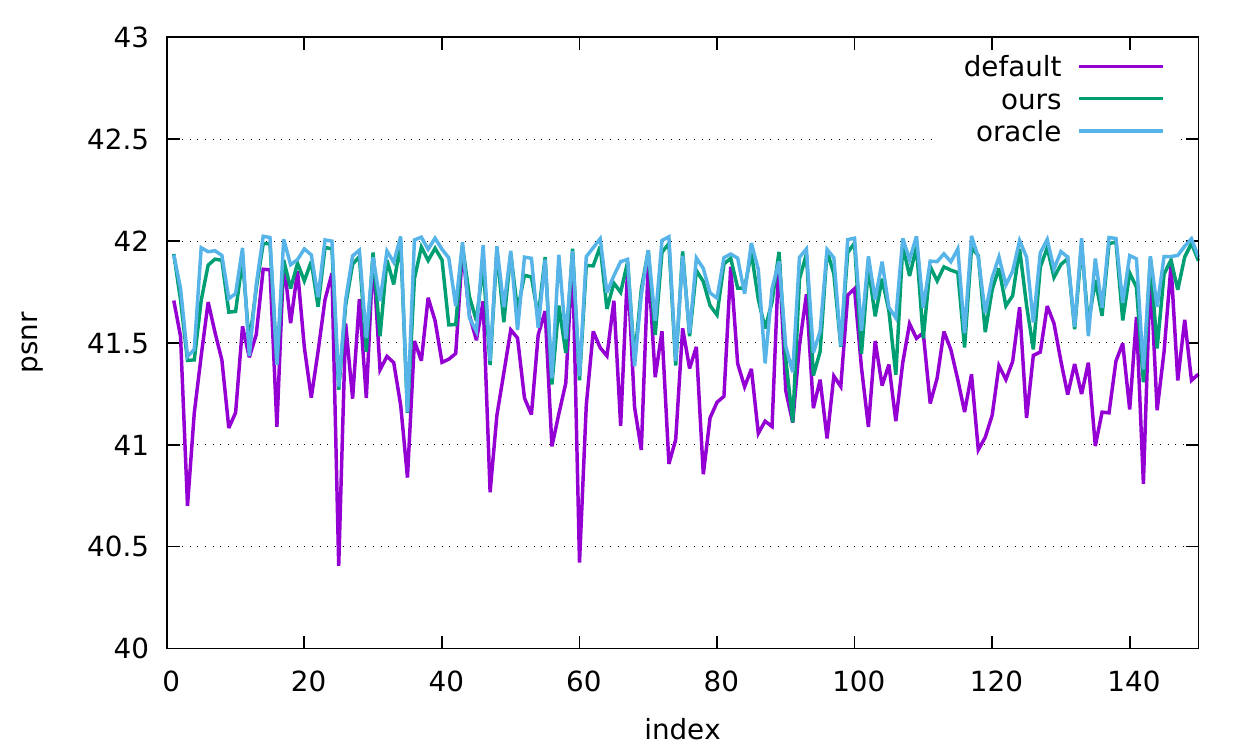}
\caption{Average PSNR}
\label{Fig:psnr}
\end{figure}
\subsection{Evaluation on multipath transmission scheme}
To evaluate the performance of the proposed path selection algorithm, two subflows are applied from sender to receiver. Two paths are available for each subflow. The experiment topology is shown in Figure \ref{Fig:multipath}. For subflow1, L1 the default path and the other path (L2 and L3) is overlay path provided by relay node. The available bandwidth in each path is randomly configured with throughput traces collected from real networks. The traced throughput dataset is publicly available at \cite{bw-trace}. The bandwidth samples are collected from mobile devices which access Internet from LTE or WiFi. 100 traces are included in this dataset and Figure \ref{Fig:trace-rate-cdf} gives the cumulative average throughput distribution. Since the trace dataset lack of transmission delay information, the transmission delay in each path is configured with delay value that is uniformly distributed between 50 milliseconds and 100 milliseconds.

Total 150 experiments are running and each simulation process lasts 400 seconds. In each test, four throughput traces are randomly chosen for this four paths. X264 \footnote{https://www.videolan.org/developers/x264.html} is used to encode video frames. The tested YUV video is KristenAndSara \footnote{https://media.xiph.org/video/derf/} (600 frames, 1280x720, 60fps) and is downloaded from \footnote{ https://blog.csdn.net/abcSunl/article/details/53841953}. Even the original video frames are captured in 60 frames per second, the video sequences are delivered to raw frame queue in 30 fps in simulation, which is enough for RTC application. Path manager makes decision to choose path for each subflow on every 1 second. When the video frames file is read to its end, its offset will be randomly chosen to align with the beginning of a frame. The available bandwidth on each routing path is traced.

The proposed path selection algorithm in multipath transmission context is compared with two other schemes ``default'' and ``oracle''. Here, ``default'' means sender exploits the default routing path in each its subflow for packet transmission and ``oracle'' means sender always chooses the routing path with maximum throughput for each subflow in each decision slot. The cumulative average bandwidth distribution of all 150 experiments is given in Figure \ref{Fig:rate-cdf}. The error bar on average throughput of all experiments in each path selection algorithm is given in Figure \ref{Fig:rate-error}. The results clearly indicate that the proposed path selection algorithm can gain higher rate in most experiments than fixed routing path transmission scheme. The ``oracle'' transmission scheme can gain the maximum profit in simulation, which is an ideal path selection mechanism. In simulation based on trace dataset, the future throughput of a path is totally known. Such algorithm is not applicable in real environment. As we argued previously, sender could not know exactly the reachable bandwidth of a routing path before sending packets into it.

At receiver side, when received packets can be reassembled into a complete frame, it will be handled over to decoder. Peak signal to noise ratio (PSNR) is calculated to evaluate picture quality after frame decoding. The average PSNR value is given in Figure \ref{Fig:psnr}. The frames involved to calculate the average PSNR in each test are about 10000. Since the proposed path selection algorithm can gain higher throughput than the default routing transmission scheme, the encoder can allocate more bit on each encoded video frames, the sender with proposed path selection algorithm can gain higher video quality as verified in Figure \ref{Fig:psnr}.
\section{Conclusion}
In this work, we combine the multipath transmission scheme with path selection for possible improvement for video telephony traffic. In real networks, the path quality is highly dynamic according to traffic load in the bottleneck link. Deploying servers in geographically distributed datacenters to form overlay network is a promising solution to improve video call quality. With the help of relay nodes in overlay network, sender can choose good paths from several candidate paths to gain maximum benefit. An online learning approach based on multi-armed bandit model is applied to select paths for subflows. Based on experiments with throughput trace collected from real networks, sender with the proposed path selection algorithm can gain higher throughput and improve video quality than the method to insist the default path transmission during the whole session.
 
As real time conversational video communication in Internet has gained quite popularity in recent years, implementing a congestion control algorithm is necessary to avoid link congestion and to promote bandwidth allocation fairness. A congestion control algorithm RTC-BBR is implemented on the multipath transmission framework, which is adapted from BBR for real time video traffic. Experiments are conducted to evaluate the performance of the proposed congestion control algorithm. Other congestion control algorithms (Cubic, BBR and GCC) are also tested to work as baselines. Results show it can achieve lower queue delay and alleviate RTT unfairness issue in BBR algorithm. Most importantly, the improved version shows less drastic throughput variation. Such property could maintain the stability of video encoder.

\bibliographystyle{elsarticle-num}
\bibliography{mp-overlay,mpvideo,ldc,shui-paper-wu,dash,support}

\end{document}